\documentclass[a4paper,twocolumn]{esapub} 
\usepackage[dvips]{graphicx}
\usepackage{natbib}
\usepackage{times}
\title{INITIAL FEATURES OF AN X-CLASS FLARE OBSERVED WITH SUMER AND TRACE }
\author{T. J. Wang}
\author{S. K. Solanki}
\author{D. E. Innes}
\author{W. Curdt}
\affil{Max-Planck-Institut f\"ur Aeronomie, D-37191 Katlenburg-Lindau,
Germany}

\newcommand{\kms}{km~s$^{-1}$}
\begin{document}

\maketitle


\begin{abstract}

A class X1.5 flare started on the solar limb at 00:43 UT on 21 April 2002, 
which was associated with a CME observed at 01:27 UT by LASCO C2.
The coordinated analyses of this flare include TRACE 195 \AA~ images  
and SUMER spectra in lines of Fe {\small XXI}, Fe {\small XII}, and C 
{\small II}. We find that:
1) The flare began with a jet seen by TRACE, which was detected by 
SUMER in the C {\small II} line as a strong brightening with blue shifts 
up to 170 \kms. At that time only weak emission was detected in Fe 
{\small XII} and Fe {\small XXI}.
2) Subsequently, a weak looplike brightening started south of the jet,
moving outwards with an average speed of about 150 \kms. The SUMER spectra 
responded this moving loop as separatingly brightenings, visible 
only in the Fe {\small XXI} line. The southwards moving component contains red-
and blue-shifted emission features and has an apparent speed of $\sim$ 
120 \kms. The absence of 
signatures in Fe {\small XII} and C {\small II} lines indicates that 
the moving weak loop seen by TRACE corresponds to the emission from very hot 
plasma, in a blend line in the 195 \AA~ bandpass due to Fe XXIV formed at 
T $>$ 10 MK.  3) The trigger mechanism of the flare and associated CME can be
interpreted in the same way as that proposed by \citet{wan02} for 
an event with similar initial features. 

\end{abstract}

\section{Introduction}

Multiple wavelength imaging of flares reveals a complex array of
features such as ejections of hot plasma in soft X-rays \citep{shi95,
ohy98} and explosive coronal waves like Moreton waves and EIT waves 
\citep{mor60, tho99}, and recent reports of X-ray waves \citep{kha02, nar02}. 
The flares with these features are found generally correlated 
with the coronal mass ejections (CMEs) \citep{nit99, kla00}. Therefore,
the information about these features are important for the understanding of
the physics of flare and CME initiation \citep{shi99, che02}. 

\begin{figure}
\centering
\includegraphics[width=1.\linewidth,height=0.6\linewidth]{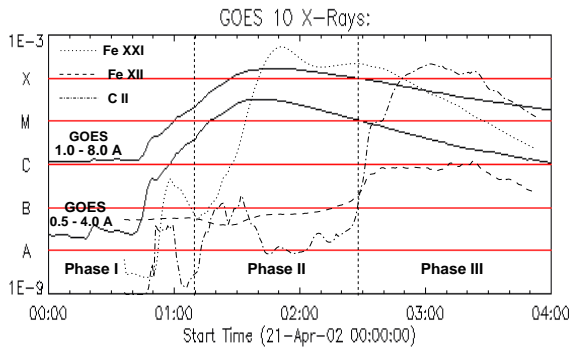}
\caption{\label{goesm}
Time profiles of the GOES flux and of line-integrated UV intensity
averaged along the SUMER slit.}
\end{figure}

The Solar Ultraviolet Measurements of Emitted Radiation (SUMER) spectrometer
aboard $SOHO$ is a powerful instrument, which can provide 
accurate Doppler shift measurements of dynamic coronal structures and  
can diagnose their temperature over a very wide range of $10^4-10^7$ K. 
Spectral measurements by SUMER of flare-associated coronal features, being
complementary to the imaging, may bring us new insight into the origin of
flares and CMEs \citep{inn01, kli02}.
In this paper, we present preliminary results derived from simultaneous spectral
and imaging observations of a class X1.5 flare that occurred just over the
southwest limb of the Sun on 2002 April 21 in active region NOAA 9906. 
The GOES soft X-ray flux began at 00:43 UT and reached a maximum at 01:51 UT 
(Fig.~\ref{goesm}). A large CME was firstly seen at 01:27 UT by the C2
telescope of the Large-Angle Spectroscopic  Coronagraph (LASCO).

\section{Observations}
\begin{figure*}[t]
\centering
\includegraphics[width=0.75\linewidth]{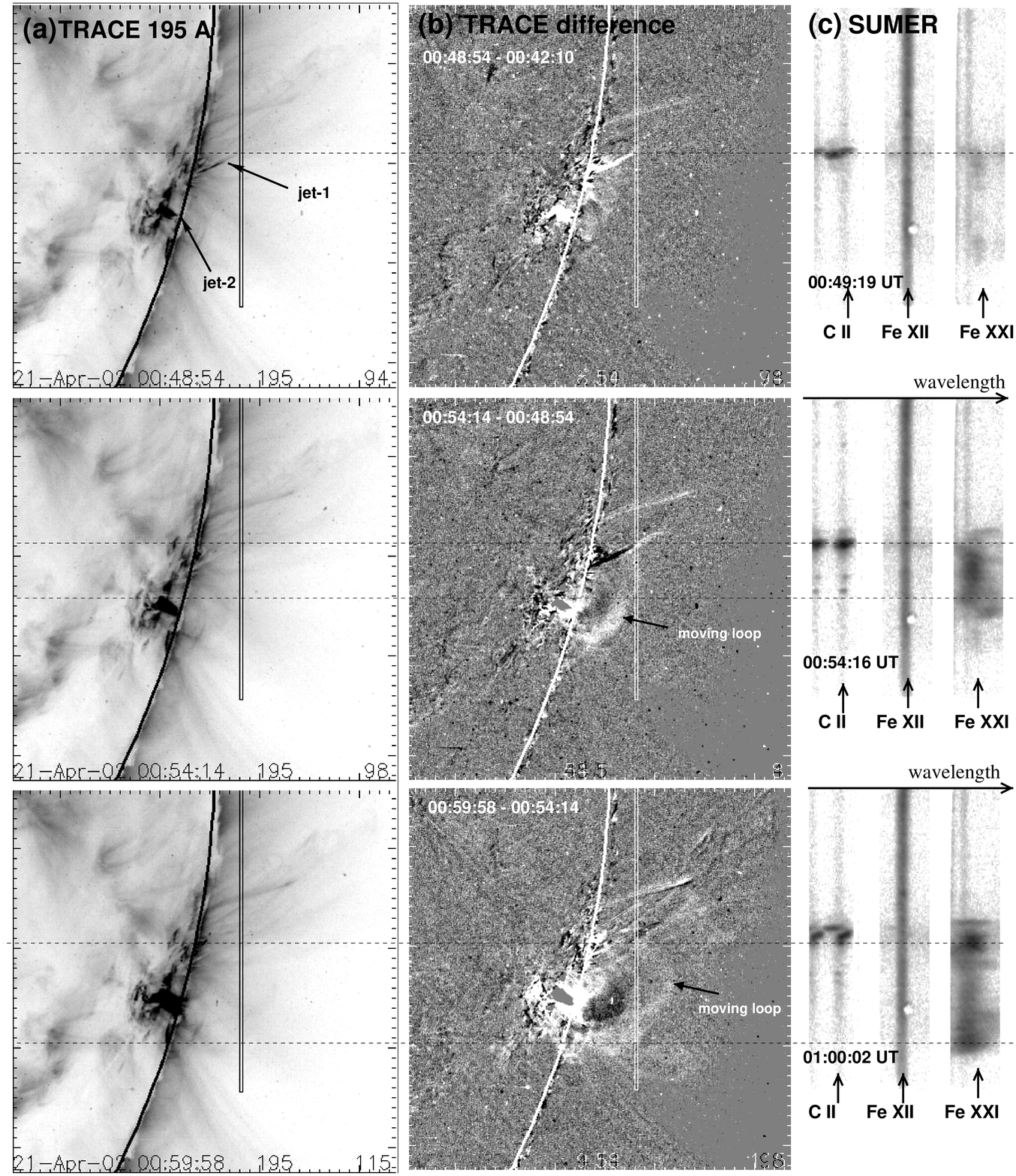}
\caption{\label{trcsm}
Association between structures seen by TRACE and SUMER at flare onset. The
position of the SUMER slit is shown in the TRACE images. (a) TRACE 195 \AA~
images, and (b) running difference images. The field of view (FOV) for TRACE 
images is $384{''}\times384{''}$. (c) SUMER spectra along the slit in three 
lines. Both TRACE images and SUMER spectra are plotted on a logarithmic 
brightness scale. (see also appended 
movies for Figures (b) and (c).)}
\end{figure*}

TRACE observed the event with a cadence of 20 s in the 195 \AA~ channel that 
contains a typical coronal line (Fe {\small XII}) and a generally much weak
flare line (Fe {\small XXIV}). SUMER recorded the spectra with a 50 s exposure 
time, and a $300{''}\times4{''}$ slit at a fixed position off the limb
(Fig.~\ref{trcsm}a). Three representative lines were recorded: a relatively
cool transition region line, C {\small II} 1335.7 \AA~ ($2\times10^4$ K),
a coronal line, Fe {\small XII} 1349.4 \AA~ ($1.6\times10^6$ K), and
a hot flare line, Fe {\small XXI} 1354.1 \AA~ ($10^7$ K). For each line,
the spectral window has a range of 2.1 \AA, thus allowing a detection of
Doppler shifts within $\sim230$ \kms. The SUMER observations started
at 00:36 UT and ended at 03:52 UT, covering the whole flare period.

\section{Results}

The EUV images show that the flare started at 00:39:29 UT as a dark jet
from the disk (see ``jet-1'' in Fig.~\ref{trcsm}a $top$). The animations of 
Figs.~\ref{trcsm}b and \ref{trcsm}c are available 
online\footnote{http://solar.physics.montana.edu/wangtj/mov}. 
As jet-1 developed, it
extended towards the slit. At 00:44:22 UT, SUMER detected the first
emission of this jet in the C {\small II} line with a blue shift of up to
170 \kms. At this moment another jetlike brightening started (see ``jet-2''
in Fig.~\ref{trcsm}a $top$). At 00:47:14 UT, jet-1 was ejected as a brightening
moving outwards at an apparent speed of 145 \kms. SUMER detected this
ejection at 00:48:29 UT in C {\small II} as a strong blue-shifted brightening
with Doppler velocity $\sim$130 \kms~, and a weak emission in the blue wing
of the Fe {\small XII} line (Fig.~\ref{trcsm}c $top$).

The TRACE observations
had a gap during 00:50$-$00:54 UT. Afterwards a weak looplike emission front
was seen moving westwards. This moving looplike feature reached the west 
boundary of the FOV at about 01:06 UT. 
Its speed increased from 100 \kms~ to more than 200 \kms~ during this period.
From 01:06 UT to 01:12 UT, the loop seemed to ungo an explosively expansion,
or opened, because the curved legs stretched rapidly. The SUMER spectra
show that in the Fe {\small XXI} line a brightening appeared at 00:49:19 UT
at the 35${''}$ south of the jet-1 and enhanced. At 00:50:58 UT, this component 
suddenly extended towards the red wing upto the window boundary, 
then the blue wing emission enhanced more strongly and moved
separately along the slit. The earlier red wing component (Doppler shift
$\sim$100 \kms) started moving southwards earlier than the blue wing component, 
so it moved ahead of the blue wing component (Fig.~\ref{trcsm}c $middle$), 
but the blue wing component caught up with the red one at 00:55:55 UT. The 
motion of multiple components suggest the presence of several expanding loops. 
The southwards moving component has an average speed of about 120 \kms, and
a maximum blue shift of $\sim$75 \kms~ at 00:58:24 UT. This south component
stopped at 01:02:31 UT 126${''}$ south of jet-1, then decayed quickly. 
While the northward moving component stopped at 00:59:13 UT just at the jet-1 
position, and kept strong emission till 01:06:39 UT, then decayed.
\begin{figure*}[t]
\centering
\includegraphics[width=0.9\linewidth]{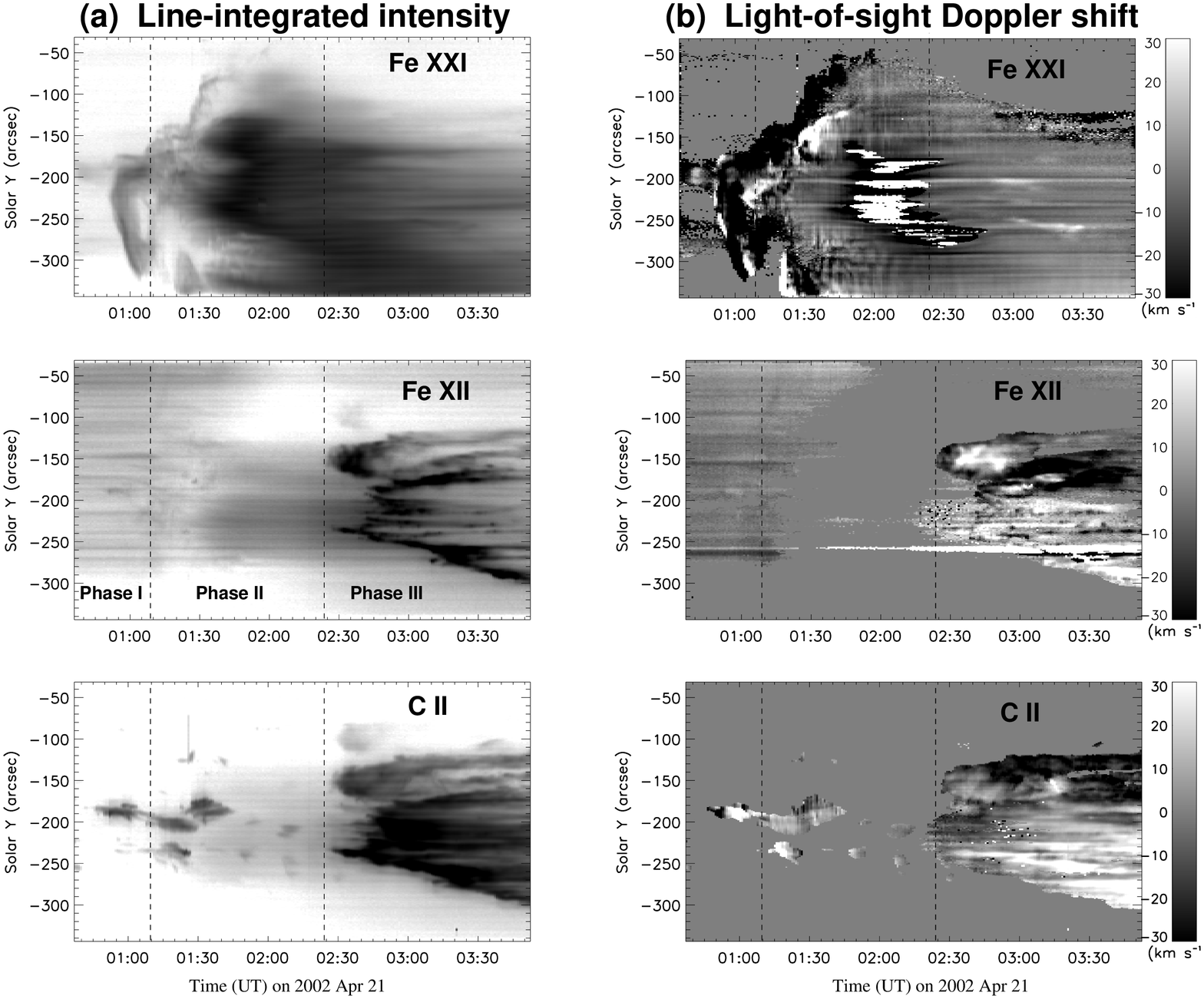}
\caption{\label{spec}
(a) Line-integrated intensity time series in three lines  on 21 April 2002,
plotted on a logarithmic scale. (b) Doppler-shift time series. In the case of
Fe {\small XXI}, the derived Doppler shifts during 01:45$-$02:45 UT at
$150{''}-300{''}$ S are not reliable due to the saturation of the detector.}
\end{figure*}

The EUV difference images show clearly that when the moving loop swept across
the slit, the SUMER spectra responded in Fe {\small XXI} line as separately
moving brightenings at the intersections of the loop with the slit 
(Fig.~\ref{trcsm}b $middle$ and $bottom$).
In addition, the SUMER spectra show that the emission of jet-1 in C {\small II}
remained at almost the same position until 01:05 UT. This implies that 
the magnetic topology related to jet-1 did not change much during this initial 
phase. Noticeably, except for weak emission in the line wing at the jet-1 
position, Fe {\small XII} showed no response to the sweeping of the EUV 
moving loop. This demonstrates that the emission of the moving
loop seen in TRACE 195 \AA~ came from the Fe {\small XXIV} line. 

The intensity time series of the SUMER spectra show the remaining jet-1
emission in C {\small II}, separately moving brightenings in Fe {\small XXI},
and no signature in Fe {\small XII} during the initial phase of 00:40$-$01:10 
UT (Phase I in Fig.~\ref{spec}a). The Doppler shift time series show 
the preflare weak red-shifted emission in Fe {\small XXI} near the 
position of jet-1 (Fig.~\ref{spec}b $top$). The Doppler shift evolution of jet-1 
shows that the shifts were first blue, then became red in C {\small II} 
(Fig.~\ref{spec}b $bottom$). 

During the main phase of 01:10$-$02:23 UT (Phase II in Fig.~\ref{spec}a),
the Fe {\small XXI} spectra show the development of red- and blue-shifted 
emission between $150{''}-240{''}$S. Then at 01:19 UT a red-shifted emission
occurred at 340${''}$S which extended northwards and shifted to the blue wing. 
At 01:24 UT, a blue-shifted emission moved from 150${''}$S to 80${''}$S at
a speed of 85 \kms. From 01:30 UT, two strong brightenings developed just 
at the sides of the jet-1 position. The C {\small II} spectra show that two
ejections occurred at 200${''}$S and 240${''}$S during 01:10$-$01:22 UT. 
From 01:25 UT to 01:45 UT, a C {\small II} ejection happened near the earlier
jet-1 position, just located at a gap between two strong brightenings in
Fe {\small XXI} (Fig.~\ref{spec}a $bottom$ and $top$). During the phase II,
except for the increase of emission in the continuum, the Fe {\small XII} 
spectra did not show any signature. 

During the decaying phase (Phase III in Fig.~\ref{spec}a), 
the simultaneous occurrence of strong emissions in Fe {\small XII} and
C {\small II} at 02:24 UT, which show a similar spatial evolution in
brightness and Doppler shift, is the most prominent feature. The brightening 
slowly moved southwards along the slit at a speed of 10 \kms. From the TRACE 
EUV images, we identify that the start of this brightening corresponded to
rising postflare arcades, which reached the slit at that time. The Fe {\small 
XXI} spectral
evolution shows no signatures in brightness and Doppler shift, in response to 
the arrival of the postflare arcades (Fig.~\ref{spec}).  But we notice 
the two brightenings in Fe {\small XII} or C {\small II} located at 
$120{''}-180{''}$ and $180{''}-270{''}$, which were divided by a strong 
emission located near the earlier jet-1. They are coincident with the
two brightenings occurring at about 01:30 UT in Fe {\small XXI}. 

Figure~\ref{goesm} shows that the soft X-ray flux increased at the time 
jet-1 was first seen in TRACE 195 \AA. The first flux pulse in C {\small II}
corresponding to the occurrence of jet-1, started little earlier than the flux
pulse in Fe {\small XXI}, which corresponds to the ejection of a hot loop.
During the main phase of the flare, the flux in soft X-ray and Fe {\small XXI} 
reached the maximum at the same time, while the flux in Fe {\small XII} and
C {\small II} were at low levels. During the decay phase, the flux
in soft X-ray and Fe {\small XXI} kept decreasing, while the rising EUV
postflare loops caused the correlated flux pulses in Fe {\small XII} and 
C {\small II} at the SUMER slit.

\section{Discussion and Conclusion}

We have analyzed the evolution of TRACE EUV images and SUMER spectra of an
X-class flare occurring at the limb. We find that the event began with an EUV
jet dominated by 10$^4$ K plasma ejection. Following this jet, a weak looplike 
brightening was seen explosively expanding outwards while being accelerated.
From the SUMER spectra, we determine the temperature of this moving emission
feature to be above 10$^7$ K, consistent with the emission from Fe {\small XXIV} 
in TRACE 195 \AA~ channel. Because the moving emission feature has a speed of
only about 150 \kms, we exclude the possibility of flare shocks. Instead,
from the curvature change, we suggest that this moving feature 
corresponds to hot coronal loop ejection, or the looplike hot plasma ejection 
seen often by SXT, associated with impulsive flares \citep{shi95}.
 
\citet{wan02} have studied a similar case of an X-class flare initiation 
associated with jets and expanding hot loops that occurred near disk center, 
and was associated with a halo CME. They found that X-ray jets happened
at a special magnetic region called a ``bald patch" (where the magnetic field 
is tangent to the photosphere) that is formed between a twisted emerging 
flux system and a transequatorial interconnecting loop system. They suggested 
that slow reconnection at the bald patch causes the jets and ejection of
the twisted flux. The initial expanding hot loops correspond to the removal
of the preheated overlaying flux due to reconnection at the bald patch. 
Their proposed scenario can also explain 
the disappearance of a soft X-ray transequatorial loop and large-scale extending
EUV dimmings, and the CME. In our case, we notice the
presence of EUV transequatorial connections between AR 9906 (the flare region)
and AR 9907 (in the north hemisphere) seen in EIT 171 \AA, and the initial
EUV jet occurring at the northern edge of the active region. Therefore, we
suggest that a similar magnetic topology and trigger mechanism as that 
proposed by \citet{wan02} can be used to interpret the flare-CME event studied 
in this paper.

We find that the brightenings in Fe {\small XXI} at phase II are coincided 
with those in Fe {\small XII} or C {\small II} at phase III.  This implies
that the rising postflare loops seen in TRACE 195 \AA~ may originate from 
the cooling
of hot loops produced via the standard flare model. And the simultaneous
occurrence of the brightenings in Fe {\small XII} and C {\small II} suggests
that the cooling time from 10$^6$ K to 10$^4$ K was very short, less than 50
s (SUMER observing cadence), or the temperature gradient in height was very
large. This may explain why the strong emission in C {\small II} lasted
longer than in Fe {\small XII}. Otherwise, if we think that they were real 
rising ``cool" loops of the plasma in temperature $10^4-10^6$ K, we need to 
explain their origin.


\begin{thebibliography}{}
\bibitem[Chen et al. (2002)]{che02}Chen, P. F., Wu, S. T., Shibata, K., 
   \& Fang, C. 2002, ApJ, 572, L99
\bibitem[Khan \& Aurass (2002)]{kha02}Khan, J. I., \& Aurass, H. 2002, A\&A, 383, 1018 
\bibitem[Klassen et al. (2000)]{kla00}Klassen, A., Aurass, H., Mann, G., 
   \& Thompson, B. J. 2000, A\&AS, 141, 357
\bibitem[Kliem et al.(2002)]{kli02} Kliem, B., Dammasch, I. E., Curdt, W.,
   \& Wilhelm, K. 2002, ApJ, 568, L61
\bibitem[Innes et al.(2001)]{inn01} Innes, D. E., Curdt, W., Schwenn,
    R., Solanki, S. K., \& Stenborg, G., 2001, ApJ, 549, L249
\bibitem[Narukage et al. (2002)]{nar02}Narukage, N., Hudson, H. S., Morimoto, S.,
    et al. 2002, ApJ, 572, L109
\bibitem[Nitta \& Akiyama (1999)]{nit99}Nitta, N. \& Akiyama, S. 1999,
    ApJ, 525, L57
\bibitem[Moreton (1960)]{mor60}Moreton, G. E. 1960, AJ, 65, 494
\bibitem[Ohyama \& Shibata (1998)]{ohy98}Ohyama, M., \& Shibata, K.
     1998, ApJ, 499, 934
\bibitem[Shibata et al. (1995)]{shi95}Shibata, K.,  1995, ApJ, 451, L83
\bibitem[Shibata (1999)]{shi99}Shibata, K., 1999, Astrophy.\& Space Sci. 
    264, 129
\bibitem[Thompson et al. (1999)]{tho99}Thompson, B. J., et al. 1999, ApJ, 
    517, L151
\bibitem[Wang et al. (2002)]{wan02}Wang, T. J., Yan, Y., Wang, J. L., Kurokawa,
 H., Shibata, K. 2002, ApJ, 572, 580
     

\end{thebibliography}
\end{document}